%% file: ifip.tex
\documentclass[10pt,conference]{IEEEtran} 
\usepackage{cite}
\usepackage{amsmath,amssymb,amsfonts}
\usepackage{algorithmic}
\usepackage{graphicx}
\usepackage{subfig}
\usepackage{textcomp}
\usepackage{xcolor}
\usepackage{tikz}
\usepackage{float}
\usepackage{grffile}
\usepackage{pdfcomment}
\usepackage{paralist} % for inline list
 \usepackage{booktabs}

\def\etal{{\em et al. }}

\definecolor{green}{HTML}{009900}

\def\BibTeX{{\rm B\kern-.05em{\sc i\kern-.025em b}\kern-.08em
    T\kern-.1667em\lower.7ex\hbox{E}\kern-.125emX}}
    
\begin{document}
%\begin{NoHyper}
\title{Impact of Traffic Characteristics on Request Aggregation in an NDN Router}
%\title{On the Effectiveness of the PIT in Reducing Upstream Demand in an NDN Router\\
%\thanks{Identify applicable funding agency here. If none, delete this.}
%}

\author{
\IEEEauthorblockN{Mahdieh Ahmadi}
\IEEEauthorblockA{%\textit{Department of Computer Engineering} \\
\textit{Sharif University of Technology}\\
Tehran, Iran \\
mahmadi@ce.sharif.edu}
\and
\IEEEauthorblockN{James Roberts}
\IEEEauthorblockA{%\textit{Department of Computer Engineering} \\
\textit{Telecom ParisTech}\\
Paris, France \\
jim.roberts@telecom-paristech.fr}
\and
\IEEEauthorblockN{Emilio Leonardi}
\IEEEauthorblockA{%\textit{Department of Computer Engineering} \\
\textit{Politecnico di Torino}\\
Turin, Italy \\
leonardi@polito.it}
\and
\IEEEauthorblockN{Ali Movaghar}
\IEEEauthorblockA{%\textit{Department of Computer Engineering} \\
\textit{Sharif University of Technology}\\
Tehran, Iran \\
movaghar@sharif.edu}
}
\maketitle

\begin{abstract}
%\\1)
%Named Data Networking (NDN)  as a promising candidate for future internet architecture incorporates collapsed forwarding technique through dedicated tables at routers called Pending Interest Tables (PIT). The PIT can decrease the bandwidth usage by aggregating the \textit{Interest} packets for the same objects. In this paper, we seek to analyze how PIT can reduce the forwarding rate of NDN on behalf of different reactive caching policies and traffic models. To investigate this metric, the behavior of caching policies under non zero download delay (non-ZDD) assumption should be studied. We propose and model for the first time an LRU variant considering pre filter and non-ZDD. We use this analytical model to provide guidelines about what traffic intensity is needed for the filter to be useful. On the other hand, quantifying the role of PIT for large catalog sizes leads us to discuss that the role of PIT in reducing forwarding rate can be negligible.
%\\2)
The paper revisits the performance evaluation of caching in a Named Data Networking (NDN) router where the content store (CS) is supplemented by a pending interest table (PIT). The PIT aggregates requests for a given content that arrive within the download delay and thus brings an additional reduction in upstream bandwidth usage beyond that due to CS hits. We extend prior work on caching with non-zero download delay (non-ZDD) by proposing a novel mathematical framework that is more easily applicable to general traffic models and by considering alternative cache insertion policies. Specifically we evaluate the use of an LRU filter to improve CS hit rate performance in this non-ZDD context. We also consider the impact of time locality in demand due to finite content lifetimes. The models are used to quantify the impact of the PIT on upstream bandwidth reduction, demonstrating notably that this is significant only for relatively small content catalogues or high average request rate per content. We further explore how the effectiveness of the filter with finite content lifetimes depends on catalogue size and traffic intensity. 
\end{abstract}

\input{introduction}
\input{relatedWork}
\input{assumptions-1}

\input{assumptions-2}

\input{analysis}

\input{evaluation}
\input{lifetime}
\input{discussion}

\input{conclusion}
\section*{Acknowledgment}
This work has  benefited from the support of NewNet@Paris, Cisco's Chair at Telecom ParisTech (\url{http://newnet.telecom-paristech.fr}). Any opinions, findings or recommendations expressed in this material are those of the authors and do not necessarily reflect the views of partners of the Chair.

\bibliographystyle{IEEEtran}
\bibliography{IEEEabrv,ifip}
%\end{NoHyper}
\end{document}

%% file: introduction.tex
\section{Introduction}
The well-known proposal for a clean-slate, Named Data Networking (NDN) architecture for the future Internet \cite{jacobson2009networking, zhang2014named} is still under active development and pre-standardization at the IRTF. 
A major feature of NDN is the systematic use of in-router, line rate caching meant to significantly reduce upstream bandwidth requirements by storing local copies of popular contents. NDN routers also perform \textit{collapsed forwarding} whereby a single content download can satisfy near simultaneous requests from multiple users. The objective of the present paper is to evaluate the effectiveness of collapsed forwarding and to understand how it depends on the traffic and popularity characteristics.

In NDN, small  chunks of content in the form of Data packets are requested by name by users who emit Interest packets. If an Interest matches a content in the router Content Store (CS), the request is a hit and the content is returned directly. If the content is absent, the request is forwarded to a Pending Interest Table (PIT). If there is no match in the PIT, the request is forwarded towards a known external source and the Interest is recorded in the PIT. If the PIT already has a matching entry, the current Interest is added to the record but not forwarded. The PIT entry is removed when the content Data packet arrives from the external source after a download delay and may then be stored in the CS. The aggregation of Interests in the PIT during the download delay thus realizes collapsed forwarding.
The PIT can be regarded as a supplementary meta cache that only stores names and not actual contents and thus potentially alleviates some serious challenges in realizing a CS of adequate capacity operating at line rate. 

To evaluate PIT effectiveness, it is clearly necessary to forego the usual assumption that downloads occur instantaneously after a request cache miss, as if there was zero download delay (ZDD). A non-ZDD assumption is required to properly account for request aggregation. We also wish to investigate the impact on performance of time locality in the request process or, more specifically, of the fact that content popularity is not constant but varies in time. Our model builds on several pieces of prior work. The  non-ZDD CS-PIT system was first analyzed by Dehghan \etal  \cite{dehghan2015analysis}. We propose an alternative approach  that is computationally more efficient when requests do not follow the usual independent reference model (IRM) but are modelled using  general renewal processes. 

We use renewal processes to model time locality illustrating its generally beneficial impact on performance compared to IRM input. The particular renewal processes considered are inspired by prior work on the analysis of ZDD systems with time locality by Garetto and co-authors in \cite{garetto2016unified}  and  \cite{garetto2015efficient}. Our analysis is applied to a CS implementing the usual Least Recently Used (LRU) replacement policy and also to a CS equipped with a filter that improves hit rates by preferentially inserting the most popular contents. The particular filter we propose is a non-ZDD variant of the 2-LRU cache considered in \cite{garetto2016unified}.
We use the analysis to perform extensive numerical evaluations, whose accuracy is confirmed by simulation, to explore the effectiveness of PIT aggregation and how this depends on critical parameters characterizing system and demand.

Our main contributions are the following:

 \begin{itemize}
 \setlength\itemsep{1em}
 \item We develop an original analytical framework to compute the hit rate and collapsed forwarding performance of  the non-ZDD CS-PIT system using LRU replacement under renewal traffic. 
  \item The analysis is extended to a 2-LRU CS-PIT system where an additional LRU meta cache filter is used to avoid caching the least popular contents. 
 \item The accuracy of the analytical framework is demonstrated by comparison with the results of simulations in an extensive series of evaluations. 
  \item These evaluations constitute an exhaustive investigation of how the effectiveness of PIT aggregation depends on download delay, CS capacity, traffic intensity and content catalogue size.
\item The impact on CS-PIT performance of finite content lifetimes (approximating varying popularity) is illustrated through results for a particular choice of renewal input.
 \end{itemize}

The rest of the paper is organized as follows. Sec.~\ref{sec:relatedWork} reviews related work. In Sec.~\ref{sec:assumptions}, we introduce the principal concepts and notations. In Sec.~\ref{sec:analysis}, we analyse LRU and 2-LRU replacement policies applied to the CS-PIT system and derive performance metrics of interest. In Sec. \ref{sec:evaluation}, we evaluate the accuracy of our analysis through extensive simulations and evaluate the performance of considered policies for contents with finite lifetime. Finally, we conclude the paper in Sec.~\ref{sec:conclusion}.

%% file: relatedWork.tex
\section{Related Work}\label{sec:relatedWork}
The literature on the modelling and analysis of caching policies is vast, as exemplified by the recent survey paper \cite{PaschosITTC18}. We limit the present discussion to papers that are most directly related to our work on the impact of PIT request aggregation in an NDN router. 

Our analysis is inspired by that proposed by Dehghan \etal in \cite{dehghan2015analysis}. Their analysis applies to  non-ZDD CS-PIT systems  under the characteristic time approximation (\cite{fagin1977asymptotic,fricker2012versatile}). 
The authors derive expressions for hit rates and request forwarding probabilities under a renewal traffic model for caching policies including LRU. 
%Our approach is different and considerably simplifies performance evaluation for renewal traffic. 
More recently, Dai \etal \cite{dai2017analysis} have considered a different implementation of LRU for the CS-PIT system where the LRU list is updated on request arrival rather than on content insertion after download. This approach makes it possible to apply to non-ZDD caches
 previous theoretical justifications of the characteristic time approximation for LRU with IRM input (\cite{fagin1977asymptotic}, \cite{fricker2012versatile}), under the  assumption that  content metadata is  never evicted from the LRU list between request arrival and download. However the assumption made in ~\cite{dai2017analysis} seems somehow artificial. This is the reason why in our  paper we prefer to restore  the more natural assumptions  made in \cite{dehghan2015analysis}.  
We apply the characteristic time approximation to non-ZDD caches, and we  validate the accuracy of this approximation by simulation.

Our work significantly extends the model of \cite{dehghan2015analysis}  by considering a more efficient cache insertion policy than simple LRU. We also develop an original, computationally efficient mathematical framework  for the non-ZDD CS-PIT system under a general renewal  traffic model, notably enabling evaluation of the impact of time locality. 

%% file: assumptions-1.tex
\section{System Assumptions}\label{sec:assumptions}
In this section, we introduce the principal concepts and notation. We discuss the assumptions used to perform the analysis in Sec.~\ref{sec:analysis} and the evaluations in Sec.~\ref{sec:evaluation}.

\subsection{CS-PIT Interplay}\label{sec:assumptions:cs}
%\todo{Considering that we explain these two components in the introduction} 
Caching policies are usually analyzed under the assumption that content downloads occur immediately after a cache miss request. In practice, in an NDN router, the delay between a CS miss and the content download can be significant and in this time, one or more subsequent requests may be aggregated in the PIT. In the following, such a request is referred to as a PIT hit while any request arriving while the content is in the CS is termed a CS hit. Let $p_{hit}^{pit}(k)$  denote the probability a request for some content $k$ is a PIT hit and $p_{hit}^{cs}(k)$ the probability it is a CS hit. Any request that is neither a PIT hit nor a CS hit is forwarded upstream so that the proportion of requests that result in a download is
\begin{equation}\label{eq:pit-hit}
{p}_{fwd}(k) = 1 - {p}_{hit}^{pit}(k)  - {p}_{hit}^{cs}(k).
\end{equation}
The round trip download delay for a forwarded request for content $k$ is assumed to be an independent random variable denoted $D_k$. In this paper, we assume that each content download request will have a response.

 \subsection{Insertion and Eviction Policies}\label{sec:assumptions:insertion}
Cache performance depends on the policies used to decide if a given content should be inserted and, if so, which other content must be evicted to make room. We limit our evaluation to two variants of the well-known LRU policy. LRU eviction is simple enough for operation at line speed and is more efficient than alternatives like FIFO or Random \cite{garetto2016unified}. The conclusions we derive on the effectiveness of the PIT are broadly the same for other policies.

We first consider the classical LRU policy where all downloaded contents are systematically inserted in the CS. We then consider a more selective insertion policy intended to improve hit rate performance. A content is only inserted on download if its name is present in a list that preferentially records popular items.  We refer to this list and its update mechanism as a `filter'. Many possible filter designs have been proposed in the literature (e.g.,  \cite{jelenkovic2005persistent}, \cite{psaras2012probcache}, \cite{elayoubi2015performance}, \cite{garetto2016unified}). The one we evaluate here is a list of a certain length that is updated using LRU. This insertion and eviction policy applied to a ZDD cache is called 2-LRU in \cite{garetto2016unified} and is shown to be an efficient solution. Its precise specification for the non-ZDD CS-PIT system is deferred to Sec.~\ref{sec:analysis:2LRU}.

%% file: assumptions-2.tex
\subsection{Content Popularity}\label{sec:assumptions:popularity}
Cache performance depends critically on how requests are spread over the population of distinct content items.  We assume here that users request items from a total population of $K$ constant size chunks. The request rate for a given chunk is determined by a popularity distribution $\{p_k\}$, $\sum_{1\le k \le K} p_k =1$, such that, if the overall request rate is $\lambda$, the request rate for content $k$ is $\lambda_k = \lambda p_k$. The content items are ordered such that $p_1 \ge p_2 \ge \cdots \ge p_k$ and we consider Zipf popularities,
\begin{equation}
p_k = k^{-\alpha} / \sum_{i=1}^K i^{-\alpha}.
\end{equation}
This choice is sufficiently general to explore the performance of the CS-PIT system under a range of popularity profiles determined by $K$ and $\alpha$. Appropriate values of these parameters are discussed in Sec.~\ref{sec:evaluation}. 
%However, we note here that an exponent  $\alpha \approx .8$ is frequently observed in practice (e.g., \cite{fricker2012impact, imbrenda2014analyzing}) leading to performance that depends significantly on $N$: for $\alpha<1$ and $N$ large, LRU hit rates for a cache of capacity $C$ depend on $C/N$ and not separately on $C$ and $N$ \cite{fagin1977asymptotic}. 
For $\alpha<1$ and $K$ large, note that ZDD LRU hit rates for a cache of capacity $C$ depend on $C/K$ and not separately on $C$ and $K$ \cite{fagin1977asymptotic}. 

Content popularities vary over time and to ignore this variability can lead to significant errors in predicting cache performance  \cite{traverso2013temporal}, \cite{OlmosKSC14}. The authors of \cite{traverso2013temporal} and \cite{OlmosKSC14} independently proposed to account for varying popularities through a so-called shot noise model. In this approach, contents appear at the instants of a stochastic process and receive requests at a rate that varies over time, eventually decreasing to zero at the end of its `lifetime'. The analysis of this model is challenging, however \cite{OlmosKSC14, leonardi2015least}. We adopt a more tractable model first proposed by Garetto \etal \cite{garetto2015efficient} .

In the model of \cite{garetto2015efficient}, contents are alternately active and inactive. An active phase has an exponentially distributed duration and corresponds to the content lifetime. During its active phases, requests for content $k$ arrive as a Poisson process of rate $\nu_k$. The inactive phase also has an exponential distribution of mean large enough that, with high probability, any cached content is evicted before the next active phase. The content thus appears in a new incarnation in each active phase. This request arrival process is known as an Interrupted Poisson process (IPP) \cite{fischer1993markov}. The overall request rate for content $k$ is 
\begin{equation}\label{eq:nu}
\lambda_k = \frac {\nu_k T_{on}}{T_{on}+T_{off}},
\end{equation}
where  $T_{on}$ and $T_{off}$ are the mean durations of active (on-period) and inactive  (off-period) phases.

\subsection{Request Process}\label{sec:assumptions:traffic}

We suppose requests for any content occur at the epochs of a \textit{stationary renewal process} \cite{samuel1975first}. 
Let $t_i$ for $i\ge 0$ be  successive request times for content $k$. The distribution of the inter-request intervals $X_i = t_{i}-t_{i-1}$ is denoted $F_k(t)$ and their density $f_k(t)$. The average request rate  is then $\lambda_k=1/\int_{0}^{\infty}(1-F_k(t))dt$. The \textit{age} of a stationary renewal process at time $t$, denoted $A_t$, is the time between $t$ and the previous request and its distribution is independent of $t$:
\begin{equation}\label{eq:age}
\mathbb P(A_t(k) < a)=\widehat{F}_{k}(a) = \lambda_k \int_{0}^{a}(1-F_k(x))dx, \textrm{for }a\ge0.
\end{equation}
The \textit{residual life} of the  interval from $t$ to the next request has the same distribution. The number of requests in an arbitrary interval of length $t$ following a request arrival (e.g., in $(t_i, t_i+t]$) is a random variable denoted $N_t$. The expectation of $N_t$  is called the \textit{renewal function} that we denote by $m_k(t)$. It satisfies the following equation,   
\begin{equation}\label{eq:counting}
m_k(t)=\widehat{F}_{k}(t)+\int_{0}^{t}m_k(t-x)f_k(x)dx.
\end{equation}

In our evaluations we consider some particular renewal processes. The simplest is the Poisson process where $F_k(t)=1-e^{-\lambda_k t}$. This choice models the so-called independent reference model (IRM) where the probability an arbitrary request is for content $k$ is independent of all previous requests and equal to $p_k$.  The IRM ignores variations in relative popularity over time and all \textit{temporal locality} between requests, i.e. the fact that if a content is requested at some instant  in time, then the probability of a request for the same content arriving in the near future increases. 

As discussed in Sec. III-C, time varying popularity can be modeled using the IPP.  This is a renewal process where intervals $(t_{i+1}-t_i)$ have a hyper-exponential distribution with two states \cite{fischer1993markov}. In Sec.~\ref{sec:evaluation} we consider a particular hyper-exponential renewal process to evaluate the accuracy of the analysis before fitting the parameters of this model to statistics derived from trace analyses.

\subsection{The Characteristic Time Approximation}\label{sec:assumptions:cta}
To evaluate CS-PIT system performance we adapt the now well-known characteristic time approximation. This approximation has become popular following its proposal by Che \etal \cite{che2002hierarchical} for evaluating LRU under the IRM, and its later analytical justification by Fricker \etal  \cite{fricker2012versatile}. It was, however, first derived as an accurate asymptotic limit by Fagin in a neglected paper from 1977 \cite{fagin1977asymptotic}. It has more recently been applied more extensively to other cache insertion and eviction policies with IRM or renewal input, notably in  \cite{garetto2016unified}.

For an LRU cache, the approximation consists in assuming a content inserted at some instant and not subsequently requested will be evicted after a deterministic characteristic time $T_C$. This represents the time for requests for $C$ distinct contents to occur where $C$ is the cache capacity. For a renewal request process, the probability an arbitrary request for content $k$ will be a hit is then
\begin{equation}\label{eq:CTA:LRU:phit}
p_{hit}(k) = F_k(T_C),
\end{equation}
while the probability the content is present in the cache at an arbitrary instant is
\begin{equation}\label{eq:CTA:LRU:pin}
p_{in}(k) = {\widehat{F}}_{k}(T_C).
\end{equation}
$T_C$ is determined on numerically solving the equation
\begin{equation}\label{eq:CTA}
C = \sum_{k=1}^K p_{in}(k).
\end{equation}

Note that the validity of the characteristic time approximation means the system can be considered as an unlimited capacity cache where contents have a constant time to live (TTL) equal to $T_C$ \cite{garetto2016unified}, \cite{fofack2014performance}. The TTL is reset to $T_C$ when the content is inserted and on every subsequent cache hit. This interpretation is used in the following analysis.

%% file: analysis.tex
\section{Performance of non-ZDD Policies}\label{sec:analysis}
We derive characteristic time approximations for the hit rate performance of a CS implemented as an LRU cache or as an LRU cache with filter, accounting for non-zero download delay.

%In this section, first we describe our design of a non-ZDD LRU policy for the CS. We exploit Che's approximation to provide a unified model for the analysis of non-ZDD policies under renewal traffic. Second, we propose a filter strategy for a non-ZDD cache which can prevent unpopular objects to be inserted into the CS. This strategy can prevent the insertion of unpopular objects in the cache therefore increasing the cache efficiency and thereby reducing the total upstream traffic. 

\subsection{LRU CS}\label{sec:analysis:lru}
\label{sec:LRU}
The LRU CS is implemented as a double linked list of pointers to stored content. Items are moved to the front of the list on insertion following a download and at the instants of subsequent requests that are hits. When a new item is inserted at the front, the last item in the list is the least recently used and is evicted. A non-ZDD LRU cache differs from classical LRU in that insertion does not occur immediately following a request miss but is deferred for a download delay $D$. Any further requests occurring in this delay are aggregated in the PIT.

 \begin{figure}[!t]
  \centering
   \subfloat[$t_E=D+T_C$]{\includegraphics[width=\columnwidth]{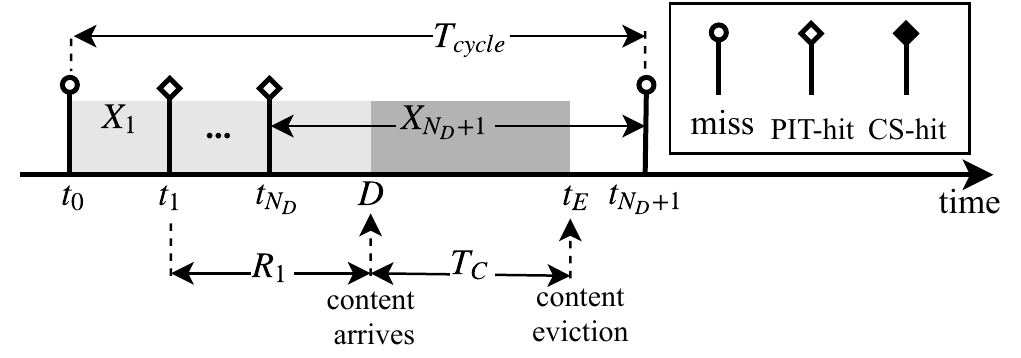}\label{fig:LRU:a}}
   \hfill
   \subfloat[$t_E=t_{N_D+N}+T_C$]{\includegraphics[width=\columnwidth]{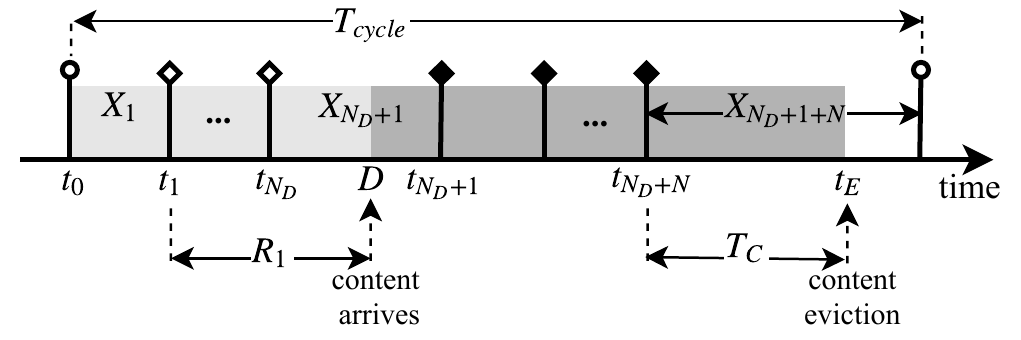}\label{fig:LRU:b}}
  \caption{Request process and CS status for a given content under non-ZDD LRU.}
  \label{fig:LRU}
  \end{figure}
  %\sum_{i=1}^{N_D+N}{X_i}
To compute hit rates, we apply the characteristic time approximation interpreting $T_C$ as the common `time to live'. We consider the sojourn of any item in the CS-PIT system that, in this interpretation, is independent of that of other items. For brevity we omit the index $k$ identifying the content in question in previously introduced notation. Fig.~\ref{fig:LRU} illustrates occupancy cycles delimited by requests that are a miss for both CS and PIT. As requests occur as a stationary renewal process, these cycles are statistically independent and hit rate performance can be derived from expected values in a typical cycle.

The cycle begins with a request miss at time $t_0$ and terminates with the next miss following content eviction at time $t_E$. Without loss of generality we set $t_0=0$. The miss at 0 triggers an upstream request (the Interest packet is forwarded towards a known source) and an initial registration in the PIT. The content is downloaded and arrives after delay $D$. Any request made between 0 and $D$ are PIT hits and are not forwarded. The number of such requests is $N_D$. 

Requests arriving between times $D$ and $t_E$ are CS hits. The number of such hits is $N \ge 0$. We have $N=0$ if the remaining inter-request interval following $D$ is greater than $T_C$ and $t_E=D+T_C$  (Fig.~\ref{fig:LRU:a}). For $N>0$, the content is evicted after the first interval that is greater than $T_C$. In this case $t_E=t_{N_D+N}+T_C$ (Fig.~\ref{fig:LRU:b}).

The performance of this system was analysed by Dehghan \etal \cite{dehghan2015analysis}. However, we found the equations derived for general renewal processes were difficult to apply in practice (due to the large computational cost) and have therefore derived a simpler, novel approximation that we now describe.

To compute $T_C$ from \eqref{eq:CTA} we need an expression for $p_{in}^{cs}$, the probability the content is in the CS at an arbitrary instant $t$. This occurs if one of the following holds:
\begin{enumerate}[(i\upshape)]
\item  the last request before $t$ was a CS hit and arrived in $[t-T_C,t)$, or  \label{event(i)}
\item  the last request before $t$ was not a CS hit, the content download occurred before $t$ and the content was not evicted before $t$. \label{event(ii)}
\end{enumerate}
Event (\ref{event(i)}) occurs when the age of the request process $A_t$ is less than $T_C$ and, from \eqref{eq:age}, has probability $\widehat{F}_k(t)$. Let $R$ be the residual download time at the arrival time of the last request before $t$. Event (\ref{event(ii)}) can then be expressed $R < A_t < R+T_C$. We deduce the expression  
 \begin{equation}\label{eq:LRU:pin}
p^{cs}_{in} = p^{cs}_{hit} \cdot\widehat{F}(T_C)+(1-p^{cs}_{hit}) \cdot\mathbb{P}(R<A_t<R+T_C).
\end{equation}
A similar argument can be applied to deduce an expression for $p^{cs}_{hit}$. In this case the situation of the content in events (\ref{event(i)}) and (\ref{event(ii)}) is considered at a request instant yielding
 \begin{equation}\label{eq:LRU:phit}
p^{cs}_{hit} = p^{cs}_{hit} \cdot {F}(T_C)+(1-p^{cs}_{hit})\cdot \mathbb{P}(R<X<R+T_C),
\end{equation}
where $X$ represents the last inter-request interval.

In the above equations, $R$ is a random variable distributed like the remaining download time of a sample request arriving at some time $t_i \in [0,D)$ with distribution $R(.)$. Fig.~\ref{fig:LRU} depicts $R_1$, the remaining download time of the first request, $R_1=D-t_1$. We proceed by first approximating the moments of $R$ and then fitting a standard distribution using moment matching. 

Let $r_n(t)$ be the sum of the $n$-th moments of the residual download times/ages  of all  requests arriving in $[0,t)$ for some constant $t \le D$: $r_n(t) = \mathbb{E}(\sum_{i=0}^{N_t} (t-t_i)^n)$, where the $t_i$ are request times and $N_t$ is the number of requests during interval $(0, t)$. To compute $r_n(t)$, we have, 
\begin{align*}
r_n(t) &= t^n + \int_{0}^{t}{r_n(t-x)dF(x)}
\\&= t^n + \int_{0}^t{(t-x)^ndm(x)},
\end{align*}
where thanks to  second equality we can rewrite the renewal equation in terms of the renewal function \cite{samuel1975first}. The moments of $R$ satisfy 
\begin{equation}\label{eq:filter-remainingDownload}
\mathbb{E}[R^n] = \frac{\mathbb{E}[r_n(D)]}{\mathbb E[ {m(D)}]+1}\approx \frac{r_n(\mathbb{E}[D])}{m(\mathbb E[D])+1}.
\end{equation}  
These moments can be used to derive a \textit{phase type} distribution that fits the distribution of $R$ arbitrarily closely \cite{whitt1982approximating}. In practice, in our numerical evaluations, it has proved sufficient to fit just the first two moments.

%In the above equations, $R$ is a random variable representing the remaining download time of a sample request that arrives before $D$ (i.e., $t_i<D$) with CDF $R(.)$. In Fig. \ref{fig:LRU}, $R_1$ is the remaining download time of first request after miss which is equal to $R_1=D-t_1$. We approximately compute the moments of $R$ as follows. First, we define $r_n(t)$ to be the sum of the $n$-th moment residual download times of all the requests that arrive before constant time $t$, where $t \le D$. $r_n(t)$ can be represented by $\mathbb{E}(\sum_{i=0}^{N_t+1} (t- t_i)^n)$, where $t_i$s are request arrival times and $N_t$ is the number of requests during interval $[0, t)$. $r_n(t)$ can be computed as  
%\begin{align*}
%r_n(t) &= t^n + \int_{0}^{t}{r_n(t-x)dF(x)}
%\\&= t^n + \int_{0}^t{(t-x)^ndm(x)},
%\end{align*}
%where the second equality transforms the renewal equation to the renewal function \cite{samuel1975first}. Then we can compute 
%\begin{equation}\label{eq:filter-remainingDownload}
%\mathbb{E}[R^n] = \frac{\mathbb{E}[r_n(D)]}{\mathbb E[ {m(D)}]+1}\approx \frac{r_n(\mathbb{E}[D])}{m(\mathbb E[D])+1},
%\end{equation}  
%where $\mathbb E[m(f, D)]$ is computed with respect to the distribution of download delay $D$. 

Solving \eqref{eq:LRU:pin} and \eqref{eq:LRU:phit}, we have,
 \begin{align}\label{eq:LRU}
p^{cs}_{in} &= \frac {\rho \widehat{F}(T_C) + \rho' (1-F(T_C))}{1-F(T_C)+\rho},\\
p^{cs}_{hit} &= \frac {\rho}{1-F(T_C)+\rho},
\end{align}
where 
\begin{align}\label{eq:LRU:rho}
\rho &= \int^{\infty}_{0}{\left(F(r+T_C)-F(r)\right)dR(r)},\notag\\
\rho' &= \int^{\infty}_{0}{\left(\widehat{F}(r+T_C)-\widehat{F}(r)\right)dR(r)}.
\end{align}
Finally, the forwarding probability is given by
\begin{equation}\label{eq:LRU:pcycle:miss}
p_{fwd}=\frac{(1-p^{cs}_{hit})}{1+\mathbb E[m(D)]},
\end{equation}
where $\mathbb E[m(D)]$ is computed with respect to the distribution of download delay $D$. 
%\JR{not necessary to recall the fixed point to compute $T_C$? not appropriate to discuss initial values...}%Using the derived $p_{in}^{cs}$, we can impose the non-ZDD LRU constraint and compute $T_C$ for this cache by Eq.~\ref{eq:CTA}. The solve the fixed point equation with less iterations, a good starting point for $T_C=C/\lambda$ where $\lambda$ is total arrival rate to the cache. Given the fact that $p_{hit}^{cs}$ is known, we can use Eq. \ref{eq:LRU:pcycle:miss} and \ref{eq:pit-hit} to calculate $p_{miss}$ and $p_{hit}^{pit}$ respectively. 
%we drive the expected value of $T_{cycle}$ using the analysis in \cite{dehghan2015analysis}(Eq. $11$) and we have
%\begin{equation}\label{eq:LRU:miss}
%{p}_{miss} = \frac{1/\lambda}{\mathbb E[T_{cycle}]}=\frac{1}{1+\mathbb E[m(fD)]+\mathbb E[N]},
%\end{equation}
%where $\mathbb E[N]$ can be calculated according to 
%\begin{equation}\label{eq:LRU:N}
%\mathbb E[N]=\frac{\int^{\infty}_{0}{\widehat F(T_C)dD(d)}}{1-F(T_C)}
%\end{equation}
%and Finally, we have 
%\begin{equation}\label{eq:LRU:p-pit}
%{p}_{hit}^{pit} = 1-{p}_{hit}^{cs} -{p}_{miss}.
%\end{equation}

\subsection{LRU CS with Filter}\label{sec:analysis:2LRU}

To improve CS hit rates we preferentially insert more popular contents, as identified by a filter placed in front of the CS. The filter consists in a double linked list of contents updated using the standard LRU policy on every \emph{request} arrival. Filter performance can thus be derived using the classical LRU characteristic time approximation: the filter hit probability for content $k$ of a filter of size $M$ is $p_{hit}^{flt}(k) = F_k(T_M)$ where characteristic time $T_M$ is such that $\sum_K \widehat {F}_k(T_M) = M$.
 
 \begin{figure}[!t]
  \centering
  \includegraphics[width=\columnwidth]{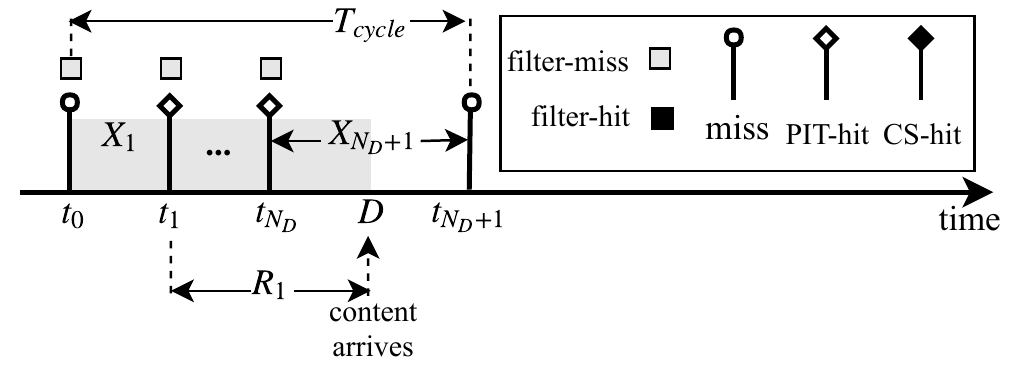}
  \caption{Request process and CS status for a given content under non-ZDD LRU with a filter.}
  \label{fig:2LRU}
  \end{figure}

A content is inserted in the CS only if on download at least one of the requests in $[0,D)$ was a filter hit. Fig.~\ref{fig:2LRU} illustrates the cycle between two forwarded requests (at $t_0$ and $t_{N_D+1}$) when the content is absent from the filter. This means the content is added to the filter at epochs $t_i$ for $0\le i \le N_D$ but always evicted before the next request at $t_{i+1}$. 

To approximate $p_{hit}^{cs}$ and $p_{fwd}$ we assume the filter and CS-PIT states are independent \cite{garetto2016unified}.  Denoting the probability of insertion by $q$ and applying the arguments of Sec.~\ref{sec:LRU} above, we deduce,
 \begin{align}
p^{cs}_{in} &= p^{cs}_{hit} \cdot \widehat{F}(T_C)+q\cdot(1-p^{cs}_{hit})\cdot \rho', \notag
\\p^{cs}_{hit} &= p^{cs}_{hit} \cdot {F}(T_C)+q\cdot(1-p^{cs}_{hit})\cdot \rho, \notag
\end{align}
where $\rho$ and $\rho'$ are given by \eqref{eq:LRU:rho}. Solving these equations gives,
\begin{align}\label{eq:2LRU}
p^{cs}_{in} &= q \cdot \frac {\rho \widehat{F}(T_C) + \rho' (1-F(T_C))}{1-F(T_C)+\rho \cdot q},\\ \label{eqq:pcsin}
p^{cs}_{hit} &= \frac {\rho \cdot q}{1-F(T_C)+\rho \cdot q}.
\end{align}

Note that these formulas would apply to any filter for which one can determine the insert probability $q$. They would apply in particular if $q$ were simply a constant probability of insertion, as envisaged in the probabilistic cache policy, called $q-$LRU in \cite{garetto2016unified}.  For the present LRU filter, the requests illustrated in Fig.~\ref{fig:2LRU} delimit independent cycles and $q=\mathbb{E}\left[1-(1-p_{hit}^{flt})^{N_D+1}\right]$. We approximate this as follows,
\begin{equation}
q \approx 1 - {(1-p_{hit}^{flt})}^{[\mathbb E[m(D)]+1]}.
\end{equation}

Recall that we have omitted dependence on the content $k$ but in fact the above probabilities are all content specific. Given expression \eqref{eqq:pcsin} for the probability a given content is present in the CS, we can determine $T_C$ numerically from \eqref{eq:CTA} and thereby, the CS  hit probabilities. The forwarding probability $p_{fwd}$ is given by \eqref{eq:LRU:pcycle:miss}, as before.

%% file: evaluation.tex
\section{Performance evaluation and insights}\label{sec:evaluation}

We first investigate the behavior of the CS-PIT system for a range of parameter settings, confirming the accuracy of our analysis by comparison with simulation results. We then use the analytical model to evaluate performance when contents have realistically long but finite lifetimes. 

\subsection{System Configuration}
We consider two instances of the request process: the Poisson process, corresponding to the IRM, and a process with 2-state hyper-exponential inter-request times: the inter-request interval for content $k$ is drawn from an exponential distribution of rate $z\lambda_k$ with probability $z/(z+1)$ and an exponential distribution of rate $\lambda_k /z$ with probability $1/(z+1)$, where $z$ is a parameter that determines the degree of time locality in the request process. We set $z=10$ to model strong correlation between requests. Observe that this process is also equivalent to an IPP where requests arrive at rate $\nu_n \approx 9\cdot \lambda_n$ in an on-period and the off-period is almost $8$ times longer than the on-period \cite[Sec. 2.3]{fischer1993markov}. We call this process `hyper10'. Note that when $z=1$, the resulting renewal process is a Poisson process and there is no time locality between requests.
\begin{table}[b]
\centering
\caption{Parameter Settings}
\begin{tabular}{llll}
\toprule
\textbf{Parameter} & \textbf{Default} & \textbf{Range} \\
\midrule
Zipf Parameter ($\alpha$) & $0.8$ & \\
Catalogue Size ($K$) & $10^6$ & $10^5$\,-\,$10^9$ \\
$[$Catalog Size/CS Capacity$]$ Ratio ($C/K$)& $10^{-3}$ & $10^{-4}$\,-\,$0.5$ \\
Download Delay ($D$) & $100$~ms & $0$\,-\,$300$~ms \\
Request Rate ($\lambda$ rqt/s)&$10^5$ & $10$\,-\,$10^6$\\
Filter Size ($M$) & $C$ \\
\bottomrule
\end{tabular}
\label{tb:parameter}
\end{table}
To validate the analysis of Sec.~\ref{sec:analysis} and to investigate system behavior, we use the parameter settings in Tab.~\ref{tb:parameter}. We have used constant download delays, the same for all contents drawn from the range reported in \cite{Singh2018CDN}.  The Zipf parameter $\alpha$ is set to 0.8 for all evaluations though we discuss the impact of alternative values in Sec.~\ref{sec:discussion}. Results are systematically presented for both LRU and 2-LRU policies.
%In this section, first we evaluate the accuracy of the analytical results for LRU and 2LRU for a particular renewal process where the intervals between requests are drawn from a two state hyper-exponential renewal process which is equivalent to an IPP. The rate of the exponential phases are defined as $\lambda_{n,1}=\lambda_n z$ and $\lambda_{n,2}=\lambda_n z^{-1}$, where $z$ is a parameter that incorporates \textit{temporal locality} between requests. We consider $z=1$ because in this case the renewal process is equal to a Poisson process and models the 'IRM'. We also consider $z=10$ which models an IPP where the requests arrive with rate $\nu_n=9.1\lambda_n$ during the active phase and $T_{on}\approx8T_{off}$ \cite[Sec. 2.3]{fischer1993markov}. We call this process 'hyper10'.

%\paragraph{Popularity}
%Zipf with $\alpha=0.8$ as default; catalogue of $10^6$ contents
%\paragraph{CS capacity}
%expressed as fraction of $N$, default $10^{-3}$
%\paragraph{Download delay} 
%default 100 ms
%\paragraph{Traffic}
%default $10^5$ requests per sec.

%\begin{table}[h]

%\vspace{2mm}
%\begin{center}
%\begin{tabular}{lll}
%parameter & default & range \\
%\hline
%popularity & Zipf(.8) & \\
%catalogue & $10^6$ & $10^5 - 10^9$ \\
%CS capacity & 1000 & $100 - 10^6$ \\
%delay ($D$) & 100 ms & $0 - 300$ ms \\
%traffic & $10^5$ rqt/s & $ 10 - 10^6$ rqt/s \\
%\hline
%\\
%\end{tabular}
%\end{center}
%\caption{Parameter settings }
%\label{tab:settings}
%\end{table}

\subsection{Performance Impacts}
Results are displayed for two performance measures: the overall CS hit probability $p_{hit}^{cs}$ and the overall forwarding probability $p_{fwd}$. The probability of PIT hit can be derived from \eqref{eq:pit-hit}.  Throughout this section, we depict the results for LRU CS with filter using the label 2-LRU. Simulation results are plotted as crosses. We have simulated a sufficient number of requests for each cross to ensure statistically stable results (up to $10^9$ requests). The plots, whose behavior is discussed below, confirm that the analytical model is generally very accurate. 
Note that 2-LRU is consistently better than LRU in all cases depicted in Figures \ref{plot:compare1} to \ref{plot:compare4}. Similarly, time locality yields consistently higher hit probabilities and lower forwarding probabilities for hyper10 traffic compared to results for the IRM. We now comment on specific impacts revealed by each set of plots.

 \subsubsection{Download Delay}\label{sec:evaluation:delay}
  \begin{figure}[!t]
  \centering
  \subfloat[LRU]{\includegraphics[width=0.5\columnwidth]{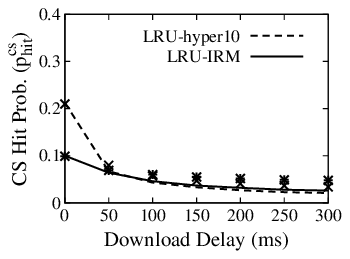}\label{plot:compare1:hit:LRU}}
   \subfloat[2LRU]{\includegraphics[width=0.5\columnwidth]{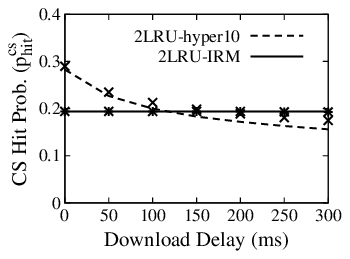}\label{plot:compare1:hit:2LRU}}\\
 %   \caption{CS hit probability}
  %\label{plot:compare1:hit}
  \subfloat[LRU]{\includegraphics[width=0.5\columnwidth]{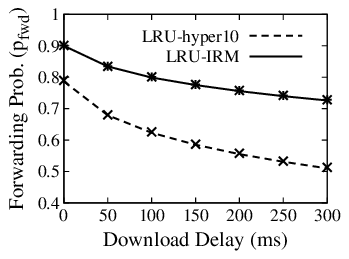}\label{plot:compare1:forward:LRU}}
   \subfloat[2LRU]{\includegraphics[width=0.5\columnwidth]{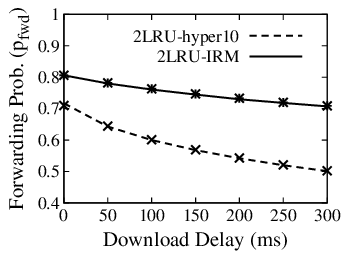}\label{plot:compare1:forward:2LRU}}
    \caption{The impact of download delay on CS hit probability and forwarding probability under LRU and 2LRU for fixed catalogue size $K=10^6$, CS capacity $C=1000$ and request rate $\lambda=10^5$ (rqt/s).}
         \label{plot:compare1}
  \end{figure}
 \figurename~\ref{plot:compare1} shows how performance depends on download delay $D$.  The CS hit probability decreases as $D$ increases but this decrease is more than compensated by an increase in PIT hits yielding a decreasing trend for the forwarding probability. The effectiveness of the PIT is clearly higher for the longer delays and may therefore bring greater benefits in more remote areas of the Internet topology. These results show the PIT plays the role of a supplementary cache and can have a significant impact on performance. The difference in $p_{fwd}$ between LRU and 2-LRU decreases as $D$ increases suggesting the PIT compensates for the absence of filter. 
%\figurename~\ref{plot:compare1} shows the CS hit probability and forwarding probability for LRU and 2LRU under IRM and hyper10 processes. The comparison of simulation and analytical results show small relative error in all the cases. The CS hit probability for both policies is increased when considering non-independent requests. Also, 2LRU policy shows more CS hit probability for all values of download delays and traffic models. 2LRU has also better resistance for large download delays. The negative slope of forwarding probability indicates that the PIT hit probability increases by download delay. The difference between forwarding probability of LRU and 2LRU gets smaller considering non-ZDD into account which indicates that PIT can aggregate the forwarded requests due to lack of filter for LRU. 

\subsubsection{CS Capacity}\label{sec:evaluation:size}
\begin{figure}[t]
  \centering
  \subfloat[IRM Traffic]{\includegraphics[width=0.5\columnwidth]{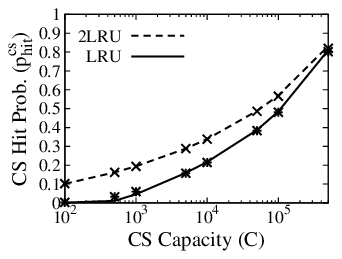}\label{plot:compare2:hit:IRM}}
   \subfloat[Hyper10 Traffic]{\includegraphics[width=0.5\columnwidth]{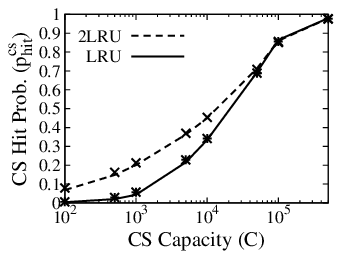}\label{plot:compare2:hit:hyper}}\\
  \subfloat[IRM Traffic]{\includegraphics[width=0.5\columnwidth]{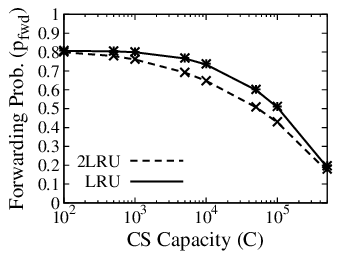}\label{plot:compare2:forward:IRM}}
   \subfloat[Hyper10 Traffic]{\includegraphics[width=0.5\columnwidth]{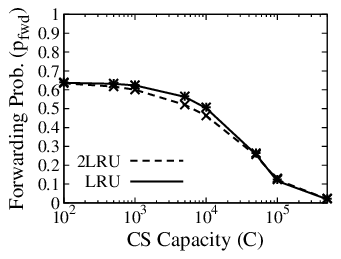}\label{plot:compare2:forward:hyper}}
    \caption{CS hit probability and forwarding probability versus CS capacity under non-ZDD LRU and non-ZDD 2LRU for fixed catalogue size $K=10^6$ and request rate $\lambda=10^5$ (rqt/s).}
  \label{plot:compare2}
  \end{figure}

\figurename~\ref{plot:compare2} illustrates the impact of CS capacity. Note that the gain in CS hits of 2-LRU over LRU is especially significant for small caches where LRU is clearly inadequate. On  the other hand, with the default download delay of $100$~ms, the reduction in $p_{hit}^{cs}$ is compensated by an increase in $p_{hit}^{pit}$ so that both policies yield nearly the same forwarding rate, especially for hyper10 traffic. This further suggests that temporal locality can also increase PIT hit probability in the same condition compared to IRM. 
%\figurename~\ref{plot:compare2} investigates the effect of the cache size on the CS hit probability and forwarding probability. For small cache sizes, 2LRU can yeild up to $500\%$ increase in the CS hit probability due to the filtering effect under IRM. Superior performance of 2LRU under hyper10 traffic (\figurename~\ref{plot:compare2:hit:hyper}) for small and moderate cache sizes indicates that the designed filter is also fast enough to adopt to fast popularity changes. For very large cache sizes, 2LRU and LRU behave almost the same especially under hyper10 traffic. In terms of forwarding probability, for small and very large cache sizes, LRU and 2LRU behave almost the same, whereas for moderate cache sizes, 2LRU can yield up to $16\%$ reduction under IRM and reduction under hyper10. Based on the above discussion, in the non-ZDD caches, the filter can improve CS hit probability for small to moderate cache sizes, but it has less improvement on the forwarding probability under correlated requests. Here, we assumed the traffic density is high enough to have reasonable Pit hit probability. In the next scenarios, we investigate if this still holds for other traffic densities.  

\subsubsection{Traffic Intensity}\label{sec:evaluation:density}
\begin{figure}[t]
  \centering
  \subfloat[IRM Traffic]{\includegraphics[width=0.5\columnwidth]{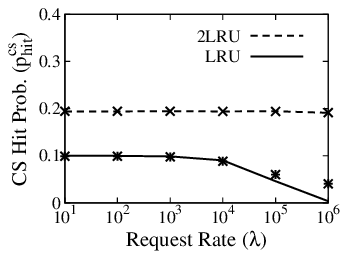}\label{plot:compare3:hit:IRM}}
   \subfloat[Hyper10 Traffic]{\includegraphics[width=0.5\columnwidth]{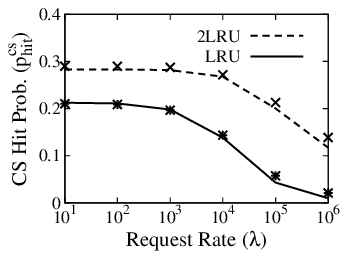}\label{plot:compare3:hit:hyper}}\\
  \subfloat[IRM Traffic]{\includegraphics[width=0.5\columnwidth]{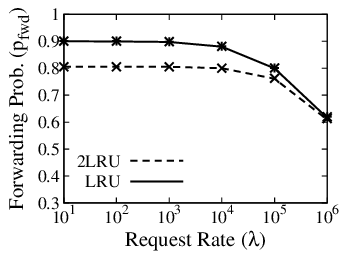}\label{plot:compare3:forward:IRM}}
   \subfloat[Hyper10 Traffic]{\includegraphics[width=0.5\columnwidth]{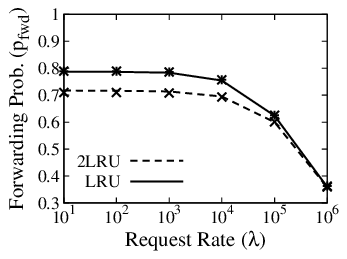}\label{plot:compare3:forward:hyper}}
    \caption{CS hit probability and forwarding probability versus request rate under non-ZDD LRU and non-ZDD 2LRU for fixed CS capacity $C= 10^{3}$ and catalogue size $K=10^6$.}
  \label{plot:compare3}
  \end{figure}

It is well known that cache performance under the ZDD assumption is independent of traffic intensity in requests/sec since it depends only on the order of requests and not on their precise timing. This insensitivity is not preserved under the present non-ZDD model. \figurename~\ref{plot:compare3:forward:IRM} for IRM input shows that the forwarding probability decreases significantly for high arrival rates thanks to an increasing probability of PIT hit. As the duration a pending request remains in the PIT is fixed at $D$ the number of aggregated requests increases in proportion to $\lambda$. Similar trends are observed for the hyper10 request process. Temporal locality of requests, however,  further accentuates the dependence of both CS and PIT performance on $\lambda$. 

 \subsubsection{Catalogue Size}
\figurename~\ref{plot:compare4} shows the impact of an increasing catalogue size. Results for very large catalogues are derived by analysis alone as simulation then becomes impractical. The catalog size varies from $10^5$ to $10^9$ while other parameters have the default settings. The case for $D=0$ (i.e., the usual ZDD assumption) is also shown for comparison. We observed in the previous analysis that high traffic densities can lead to a decrease of the CS hit probability as more requests miss the CS during download time. In \figurename~\ref{plot:compare4}, the per-content traffic intensity decreases as the catalogue size grows leading therefore to a CS hit probability that increases and tends to the ZDD value. \figurename~\ref{plot:compare4:forward:IRM} suggests that the PIT hit probability under IRM input is negligible for large catalogues,  i.e., under low per-content traffic intensity. For hyper10 traffic, on the other hand, request time locality means PIT aggregation remains effective for bigger catalogues and the non-ZDD model is necessary to accurately predict performance.    

%% file: lifetime.tex
\subsection{Impact of Finite Lifetime}
We now complement the extensive evaluation scenarios of the previous section  using a more realistic model of popularity variation. The hyper-z model can artificially model temporal locality but hardly represents realistic variations since high activity periods, representing finite lifetimes, have the same mean number of requests so that content lifetime durations are inversely proportional to popularity. In this section we assume lifetimes have a given average duration. For illustration purposes, we set the same lifetime for all contents though the model would allow content specific durations.
 \begin{figure}[t]
  \centering
  \subfloat[IRM Traffic]{\includegraphics[width=0.5\columnwidth]{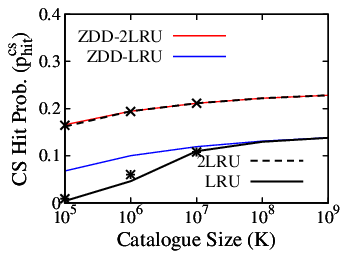}\label{plot:compare4:hit:IRM}}
   \subfloat[Hyper10 Traffic]{\includegraphics[width=0.5\columnwidth]{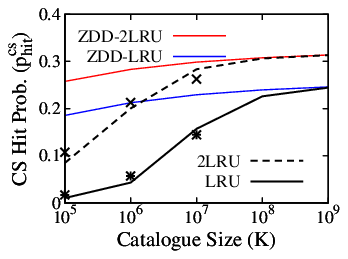}\label{plot:compare4:hit:hyper}}\\
  \subfloat[IRM Traffic]{\includegraphics[width=0.5\columnwidth]{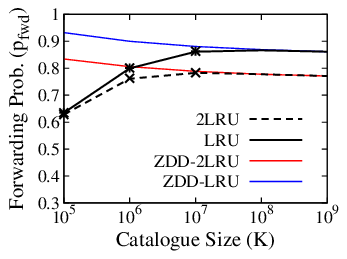}\label{plot:compare4:forward:IRM}}
   \subfloat[Hyper10 Traffic]{\includegraphics[width=0.5\columnwidth]{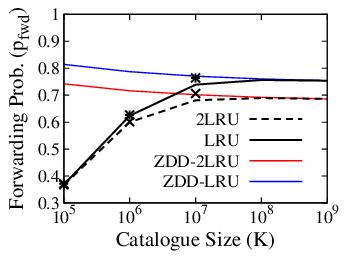}\label{plot:compare4:forward:hyper}}
    \caption{CS hit probability and forwarding probability versus catalogue size under LRU and 2LRU (ZDD and non-ZDD) for fixed $C/K$ ratio $=10^{-3}$ and request rate $\lambda=10^5$ (rqt/s).}
  \label{plot:compare4}
  \end{figure}

Measurements reported in the literature show that the average lifetime of the most dynamic fraction of video on demand content is around $2$ days (see \cite{traverso2013temporal}). 
%As it is clearly not practical to simulate a system with such a wide timescale for lifetimes ($> 1$ day) for an arbitrary request rate, we rely here on analytical results to be able to explore wider range of possible values.   
%As it is clearly not practical to simulate a system with such a wide timescale between lifetime, for arbitrary request rate...
As it is clearly not practical to simulate a system with such a wide difference in timescales between lifetimes ($> 1$ day) and download delays ($< 1$ sec), we rely here on analytical results. As explained in Sec.~\ref{sec:assumptions:traffic}, we model time varying popularity using IPP renewal processes. The lifetime is identified with an exponential on-period of mean duration  $T_{on}$ while the exponential off-period is of mean duration $T_{off}=9\cdot T_{on}$. $T_{off}$ must be large enough that it is considerably larger than the characteristic time $T_C$ so that each on-period appears as a new content with respect to CS and PIT states.

We take parameter values from the paper \cite{garetto2015efficient} where the IPP model was first proposed. $T_{on}$ and catalogue size $K$ are set so that the rate at which `new' contents occur, denoted $\gamma$, and the mean number of active contents are fixed. Thus $K=10 \gamma T_{on}$. We set $\gamma=5\cdot 10^4$ contents per day, the value reported in \cite{garetto2015efficient}, and select $T_{on}$ from 1, 7 and 30 days to explore a range of scenarios. Cache size is set to $C=0.01K$ here and other parameters take default values from Tab. \ref{tb:parameter}. % The popularity distribution is Zipf with parameter 0.8. but this is default

\begin{figure}[t]
  \centering
  \subfloat[$T_{on}=1$ day]{
  \includegraphics[width=0.5\columnwidth]{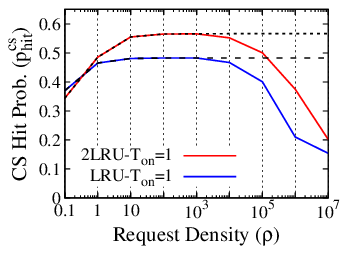}
  \includegraphics[width=0.5\columnwidth]{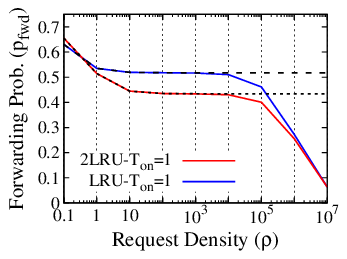}
\label{plot:lifetime3:1}}\\
\subfloat[$T_{on}=7$ days]{
\includegraphics[width=0.5\columnwidth]{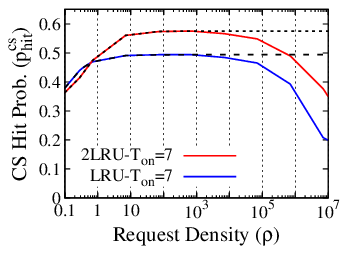}
\includegraphics[width=0.5\columnwidth]{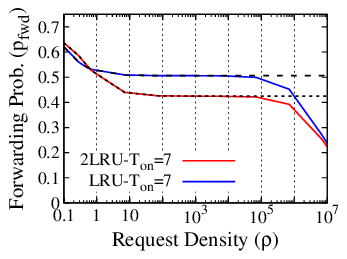}\label{plot:lifetime3:7}}      
 \caption{CS hit probability and forwarding probability versus request density for different $T_{on}$ durations under non-ZDD LRU and non-ZDD 2LRU when CS capacity is fixed to $C=0.01K$. Black dashed lines represent the case for $D = 0$.}
 \label{plot:lifetime3}
\end{figure}

%The extensive scenarios tested in the previous section showed the accuracy of analytical model in predicting our metrics of interest. In this section, we use the analytical model to predict the metrics under time varying popularity.  To set realistic traffic parameters we resort on the statistic provided  in \cite{traverso2013temporal} and \cite{garetto2015efficient}.
%Available measurements show that  the  average lifetime for the most  dynamic fraction of  contents in a Video on Demand (VoD) 
%context is around $2$ days \cite{traverso2013temporal}. As explained in Sec.~\ref{sec:assumptions:traffic}, in this paper we model  varying popularity by stationary renewal processes, 
%identifying  content lifetime with  $T_{on}$.  We also consider an extreme case where contents exhibith very fast dynamic, by setting $T_{on}=1$. 
%% 
%We set $T_{off}=9\cdot T_{on}$ which is large enough to be sure about content eviction after the on-period in the considered scenarios.
% As in \cite{garetto2015efficient}, we set the catalog size equal to $K=\gamma \cdot T_{on} \cdot \frac{T_{on}+T_{off}}{T_{on}}$ where $\gamma$ is the arrival rate of new contents per day in the SNM model and $\gamma=5\cdot10^4$. 
%We assume the request rate for contents in the on-period ($\nu_k$) follows Zipf with $\alpha=0.8$ for all the evaluations where $\nu_k$ is related to $\lambda_k$ according to \eqref{eq:nu}.   
\figurename~\ref{plot:lifetime3}  plots $p_{hi}^{cs}$ and $p_{fwd}$ as functions of a measure of request density, denoted $\rho$, for $T_{on}=1$ day and $T_{on}=7$ days. Request density is defined as the expected total number of requests occurring in a time-window  of duration equal to a content lifetime: 
$$\rho=\sum \nu_k/K T_{on} =\lambda/\gamma.$$

Cases (a) and (b) behave similarly for small $\rho$. The CS hit probability is small and, moreover, the 2LRU policy becomes less effective than LRU. This is explained by the low reactivity of these policies at low density: the first request and first two requests in an on-period are necessarily misses for LRU and 2LRU, respectively, and the relative impact of these systematic misses is significant when $\rho<10$.

The behavior of both cases  is qualitatively similar for large $\rho$ though quantitatively different. This reflects the beneficial impact of the PIT in aggregating requests when there is a large number of requests per lifetime. The PIT is more effective when the ratio of download delay to lifetime is larger and/or the requests show more time locality, i.e., for $T_{on}=1$. The performance of the CS-PIT system has the same behavior as a ZDD cache for low densities, $\rho < 10^4$, since collapsed forwarding then becomes negligible.

\begin{figure}[t]
  \centering
  \subfloat[CS Hit Ptobability]{\includegraphics[width=0.52\columnwidth]{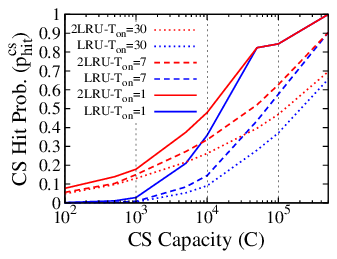}\label{plot:lifetime2:tota:hit}}
   \subfloat[Forwarding Probability]{\includegraphics[width=0.52\columnwidth]{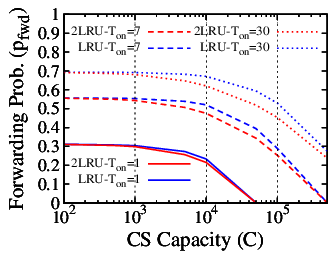}\label{plot:lifetime2:total:forwarding}}
    \caption{CS hit probability and forwarding probability versus CS capacity for different $T_{on}$ durations under non-ZDD LRU and non-ZDD 2LRU when request density is fixed to $\rho=10^6$.}
  \label{plot:lifetime2:total}
  \end{figure}

\figurename~\ref{plot:lifetime2:total} plots $p_{hit}^{cs}$ and $p_{fwd}$ as functions of CS capacity for different values of the average on-period duration. We set the request density to $\rho=10^6$ in this scenario. The results show that CS hit probability is roughly inversely proportional to the length of on-period, i.e., content lifetime. On the other hand PIT effectiveness increases when the on-period duration is reduced for a fixed $\rho$ value.

%% file: discussion.tex
\subsection{Discussion} \label{sec:discussion}
The results of this section show that the effectiveness of PIT aggregation varies widely depending on the chosen scenario. It is useful therefore to discuss observed behavior in the light of known demand and network characteristics. 

The PIT is very effective in reducing forwarding rates when demand is high and concentrated on a relatively small catalogue of contents (cf. \figurename~\ref{plot:compare3} and \figurename~\ref{plot:compare4}). This may be true for certain NDN deployments but perhaps not so for core routers performing content retrieval from the entire Internet. Known statistics on different types of content, like the web or YouTube videos, suggest catalogues approaching the petabyte in total volume \cite{fricker2012impact}, while traces from real traffic observations reveal volumes of at least several tens of terabytes \cite{imbrenda2014analyzing}. Converted to NDN chunks (close in size to IP packets) suggests catalogues $K$ in excess of $10^{10}$. On the other hand, demand in a core NDN router with multiple 10 Gb/s links might generate $O(10^6)$ requests per second at peak times. The request rate per content item ($O(10^{-4}))$ is still rather low for the PIT to be effective (in \figurename~\ref{plot:compare3}, the same relative per content request rate would occur at $\lambda=100$).  The scenario would clearly be more favorable with a more skewed popularity distribution (e.g., Zipf(1)) though most observations suggest this is not very likely (e.g., \cite{imbrenda2014analyzing}).

The plots in Figures \ref{plot:compare1}  to \ref{plot:compare4} demonstrate the generally positive impact on hit rate performance of time locality. On the other hand, \figurename~\ref{plot:lifetime3} shows that finite lifetimes can significantly reduce the effectiveness of reactive caching policies like LRU and 2LRU when demand is relatively low.  The critical parameter is the expected number of requests in a content lifetime. If this is small, in an edge router delivering content from a large catalogue say, it may be necessary to perform proactive caching (i.e., to push the most popular contents to the CS) in order to significantly reduce the forwarding rate.

%% file: conclusion.tex
\section{Conclusion}\label{sec:conclusion}
The characteristic time based analytical framework developed in this paper is both versatile and accurate. We have modelled the CS-PIT system with non-zero download delay applying LRU and 2-LRU cache replacement policies under general renewal request processes. Analytical results, whose accuracy is confirmed by simulations, enable an appraisal of the effectiveness of the PIT in reducing network traffic through the use of collapsed forwarding.

The effectiveness of the PIT naturally increases with the duration of the download delay. The more complex 2-LRU replacement policy gives higher CS hit rates than simple LRU in all cases but this advantage is mitigated in non-ZDD scenarios where the PIT is a meta cache that, like the filter, tends to improve the performance of the most popular contents.

The PIT is most effective when demand per content is relatively high such that several requests often occur in a download delay. This happens when overall demand is high, as in a core router, but only if this demand is not spread over a very large content catalogue. If demand is low, as in an access node or an enterprise router, and nevertheless spread over a large catalogue, the PIT is hardly effective in realizing collapsed forwarding. 

When contents have a finite lifetime during which they are popular and receive requests (an approximate model of popularity variation), the above remarks on PIT effectiveness still apply. In addition we observed that both reactive cache policies, LRU and 2-LRU, can be ineffective when the expected number of requests per content per lifetime is small. Whenever this case arises in practice it appears necessary to implement a placement policy where the most popular contents are proactively pushed to the CS.

% 
%
%To evaluate the effectiveness of PIT in reducing upstream bandwidth in an NDN router, we have developed an analytical framework that extends\cite{dehghan2015analysis}.
%Our approach allows us to analyze the performance of non-ZDD LRU caching systems under renewal traffic at limited computational cost.   Moreover  our approach can be applied to more advanced caching systems  implementing more sophisticated  insertion policies.
% The accuracy of the proposed framework has been extensively tested against simulations.  Our analytical tool permits us to evaluate the effectiveness of the PIT as well as  
% the impact of  sophisticated  insertion policies in dynamical scenarios in which  content popularities varies with time, i.e. contents exhibit temporal locality.
% Our  results show that in general   the presence of a filter can improve the performance of caching system (under non-ZDD assumptions), however  in some corner cases
%  (highly dynamical scenarios with very low density) the presence of a filter can worsen the performance 
%
%PIT, in general, is  effective  when the request rate is pretty high and the   CS size is  moderate, so that misses of highly requested content are frequent.
%In such conditions, PIT acts as  a supplementary cache which  reduces the forwarding rate
%compensating,  at least in part,  caches deficiencies.
% Therefore,  the beneficial effect of the presence of an insertion  filter  on forwarding rate is stronger for moderate traffic densities and/or 
%  small CS capacities (i,.e., when the PIT effectiveness is low). 